\documentclass{tlp}

\author[A. Dovier and V. Santos Costa]{
Agostino Dovier \\
Dip.~di Matematica e Informatica, Univ.~di Udine, Italy\\
\email{agostino.dovier@uniud.it}
\and
V\'{\i}tor Santos Costa\\
CRACS INESC-TEC and Dep.~de Ci\^encia de Computadores, Univ.~do Porto, Portugal\\
\email{vsc@dcc.fc.up.pt}
}

\title[Introduction to the 28th ICLP Special Issue]
{Introduction to the 28th International\\
Conference on Logic Programming\\
Special Issue}

\volume{12 (4--5)}

\pagerange{421--426}

\pubyear{2012}

\jdate{July 2012}

\doi{S1471068412000300}

\newcommand{\tocTitle}[2]{\ \\ [-2.2mm] #1\\ [1mm]}
\newcommand{\tocAuthors}[1]{\hspace*{4mm}%
\begin{minipage}{0.9\textwidth}
\emph{#1}\\ [1.6mm]
\end{minipage}}

\begin{document}
\maketitle


We are proud to introduce this special issue of the Journal of Theory
and Practice of Logic Programming (TPLP), dedicated to the full papers
accepted for the 28th International Conference on Logic Programming
(ICLP).  The ICLP meetings started in Marseille in 1982 and since then
constitute the main venue for presenting and discussing work in the
area of logic programming.

We contributed to ICLP for the first time in 1991. The first guest-editor
had a paper on logic programming with sets, and the second had two
papers on the parallel implementation of the Andorra model.  Since
then, we continued pursuing research in this exciting area and ICLP
has always been the major venue for our work.
Thus, when the ALP EC committee kindly invited us for chairing the
2012 edition we were delighted to accept.

We particularly appreciate the honor and responsability of organising
ICLP in Budapest. Hungary has had a central role both in
implementation and in the application of logic programming. Indeed,
the role of Hungary in general in Computer Science is widely
recognized, and organizing this meeting in the town of John von
Neumann, one of the ``talent-scouts'' of Turing, in the centenary of
the birth of the latter, is just another reason for justifying the fact
that the fascinating Budapest is the unique town to host ICLP twice.

Publishing the ICLP full papers as a special issue is a joint
initiative taken by the Association for Logic Programming and by
Cambridge University Press. The goal is to achieve fast journal
publication of the highest quality papers from the logic programming
community, by selecting the best ICLP submissions before the meeting.
This approach benefits the authors, by facilitating journal
publication, and benefits the community, by allowing researchers to
access high quality journal papers on the more recent and important
results in the field. Quality is ensured by a two-step refereeing
process, and by an active and very much participating program
committee. The approach was first experimented in 2010, and has had
favorable feedback since.

This year, ICLP sought contributions in all areas of logic
programming, including but not restricted to the following
areas. \emph{Theory:} Semantic Foundations, Formalisms, Non-
monotonic Reasoning, Knowledge Representation; \emph{Implementation:}
Compilation, Memory Management, Virtual Machines, Parallelism;
\emph{Environments:} Program Analysis, Transformation, Validation,
Verification, Debugging, Profiling, Testing; \emph{Language Issues:}
Concurrency, Objects, Coordination, Mobility, Higher Order, Types,
Modes, Assertions, Programming Techniques; \emph{Related Paradigms:}
Abductive Logic Programming, Inductive Logic Programming, Constraint
Logic Programming, Answer-Set Programming; \emph{Applications:}
Databases, Data Integration and Federation, Software Engineering,
Natural Language Processing, Web and Semantic Web, Agents, Artificial
Intelligence, Bioinformatics.


In response to the call for papers we received 102 abstracts, 90 of which
remained as complete submissions.  Of these, 81 were submitted as full
papers and 9 as technical communications. Each paper was reviewed by at least
three anonymous program committee members, selected by the program
chairs. Sub-reviewers were allowed.  After discussion, involving the
whole program committee, 10 submitted papers were considered as
deserving of TPLP publication with minor changes. The authors of other
10 submitted paper were asked to address more serious concerns, mostly
regarding presentation improvements or more complete experimental
validation.  37 papers instead have been judged to deserve a slot for
a short presentation at the Meeting and a ``technical communication''
publication in the Volume 17 of the Leibniz International Proceedings in
Informatics (LIPIcs) series.


The whole set of accepted papers includes 36 technical papers, 12
application papers, 5 system and tool papers, and 4 papers submitted
directly as technical communications.

The Conference program was honored to include contributions from three
keynote speakers and from a tutorialist. Two invited speakers come
from industry, namely Ferenc Darvas from \emph{ThalesNano} (a Budapest
company specialized in developing and providing microscale flow
instruments for chemistry), and Mike Elston from \emph{SecuritEase}
(an Australian company developing stock brokering tools).  Moreover,
Jan Wielemaker, of the VU University Amsterdam, presented an history
of the first 25 years of SWI Prolog, one of the major (and free)
Prolog releases. Tutorialist Viviana Mascardi from University of
Genova (Italy) introduced us to the hot topic of ``Logic-based Agents and
the Semantic Web''.

\medskip

The first ICLP Conference was organized 30 years to this year, in Marseille.
During those 30 years, ICLP has been a major venue in Computer Science.
In order to acknowledge some of the major contributions that have been fundamental to the success of LP as a field, the  ALP executive committee
decided that ICLP should recognize the most influential papers presented in the ICLP and ILPS conferences
(ILPS was another major meeting in logic programming, organized until 1998),
that, 10 and 20 years onwards, have been shown
to be a major influence in the field.  As program co-chairs of
ICLP2012, we were the first to be charged with this delicate task. We
included papers from ICLP 1992 and ILPS 1992 , 20 years onwards,
and of ICLP 2002, 10 years onwards.  Our procedure was to use
biblio-metric information in a first stage, and to use our own
personal criteria in a second stage, if necessary. Given that this is
the first time this award was given we also considered 1991, and 2001
papers.  Although there are an impressive number of excellent papers
in 1991 and 1992, one paper emerges with an outstanding record of
roughly 600 citations. Further, the paper clearly has a very major
influence in the field. The paper is

\begin{itemize}
\item Michael Gelfond and Vladimir Lifschitz: Representing Actions in Extended Logic Programming. JICSLP 1992: 559-573
\end{itemize}

The 10 years onward analysis again produced a group of excellent
papers (as expected, the number of citations was strictly less than for
20 years old papers). In this case choosing the winner in a very short
list was more difficult. Ackowledging their influence over the very
active field of Web Databases and Semantic Web, our selection went to:

\begin{itemize}
\item Fran\c{c}ois Bry and Sebastian Schaffert: Towards a Declarative Query and Transformation Language for XML and Semistructured Data: Simulation Unification. ICLP 2002: 255-270
\end{itemize}

We therefore invited these authors for an invited talk in a special
session at the meeting. We would like to remark that in 2004
three papers were prized as ``Most Influential Paper in 20 Years Award from the Association for Logic Programming''.
The award went (again) to Gelfond-Lifschitz for their 1988 ICLP/SLP paper on stable model semantics,
to Jaffar-Lassez for their POPL 1987 paper on Constraint Logic Programming, and
to Saraswat, Rinard, and Panangaden for their POPL 1991 paper on Concurrent Constraint Programming.

\medskip

Together, the journal special issue and the LIPIcs volume of short technical
communications constitute the proceedings of ICLP 2012.
The list of the 20 accepted full papers appearing 
in this special issue follows:

\medskip

\tocTitle{Disjunctive Datalog with Existential
Quantifiers: Semantics, Decidability, and Complexity Issues}{00}
\tocAuthors{Mario Alviano, Wolfgang Faber, Nicola Leone, and Marco Manna}

\tocTitle{Towards Multi-Threaded Local Tabling Using a Common Table Space}{00}
\tocAuthors{Miguel Areias and Ricardo Rocha}

\tocTitle{Module Theorem for the General Theory of Stable Models}{00}
\tocAuthors{Joseph Babb and Joohyung Lee}

\tocTitle{Typed Answer Set Programming Lambda Calculus and
  Corresponding Inverse Lambda Algorithms}{00}
\tocAuthors{Chitta Baral, Juraj Dzifcak, Marcos Gonzalez, and Aaron
  Gottesman}

 \tocTitle{ D-FLAT: Declarative Problem Solving Using Tree
Decompositions and Answer-Set Programming}{00}
\tocAuthors{Bernhard Bliem, Michael Morak, and Stefan Woltran}

\tocTitle{An Improved Proof-Theoretic Compilation of Logic Programs}{00}
\tocAuthors{Iliano Cervesato}

\tocTitle{Annotating Answer-Set
Programs in LANA}{00}
\tocAuthors{Marina De Vos, Doga Gizem Kisa, Johannes Oetsch, J\"org P\"uhrer, and Hans Tompits}

\tocTitle{SMCHR: Satisfiability Modulo Constraint Handling Rules}{00}
\tocAuthors{Gregory Duck}

\tocTitle{Conflict-driven ASP Solving with
External Sources}{00}
\tocAuthors{Thomas Eiter, Michael Fink, Thomas Krennwallner, and Christoph Redl}

\tocTitle{Multi-threaded ASP Solving with clasp}{00}
\tocAuthors{Martin Gebser, Benjamin Kaufmann, and Torsten Schaub}

\tocTitle{Model Checking with Probabilistic Tabled Logic
Programming}{00}
\tocAuthors{Andrey Gorlin, C. R. Ramakrishnan, and Scott Smolka}

\tocTitle{Diagrammatic confluence for Constraint Handling Rules}{00}
\tocAuthors{R\'emy Haemmerl\'e}

\tocTitle{Inference in Probabilistic Logic Programs with Continuous Random Variables}{00}
\tocAuthors{Muhammad Islam, C.R. Ramakrishnan, and I.V. Ramakrishnan}

\tocTitle{Relational Theories with Null Values and
Non-Herbrand Stable Models}{00}
\tocAuthors{Vladimir Lifschitz, Karl Pichotta, and Fangkai Yang}

\tocTitle{The Relative Expressiveness of Defeasible Logics}{00}
\tocAuthors{Michael Maher}

\tocTitle{Compiling Finite Domain Constraints to SAT with BEE}{00}
\tocAuthors{Amit Metodi and Michael Codish}

\tocTitle{Lightweight Compilation of (C)LP
to JavaScript}{00}
\tocAuthors{Jose F. Morales, R\'emy Haemmerl\'e, Manuel Carro, and Manuel Hermenegildo}

\tocTitle{ASP modulo CSP: The clingcon system}{00}
\tocAuthors{Max Ostrowski and Torsten Schaub}

\tocTitle{Annotation of Logic Programs for Independent AND-Parallelism by Partial Evaluation}{00}
\tocAuthors{German Vidal}

\tocTitle{Efficient Tabling of Structured Data with Enhanced Hash-Consing}{00}
\tocAuthors{Neng-Fa Zhou and Christian Theil Have}

The technical communications, a short paper, and the contributions to
the doctoral consortium are published on-line through the Dagstuhl
Research Online Publication Server (DROPS), as DOI
10.4230/LIPIcs.ICLP.2012.42. A listing of these papers is
reported in the electronic appendix to this preface.

\medskip

We would like to take this opportunity to acknowledge and thank the
other ICLP organisers. Without their work and support this event would
not have been possible.  We would like to start with the General chair
P\'eter Szeredi (Budapest Univ. of Technology and Economics), and all
the organizing chairs, namely the Workshop Chair Mats Carlsson (SICS,
Sweden), the Doctoral Consortium Chairs Marco Gavanelli (Univ. of
Ferrara) and Stefan Woltran (Vienna University of Technology), the
Prolog Programming Contest Chair Tom Schrijvers (Universiteit Gent),
the Publicity Chair Gergely Luk\'acsy (Cisco Systems Inc.), and the
Web Manager: J\'anos Csorba (Budapest Univ. of Technology and
Economics).  Thanks also to Alessandro Dal Pal\`u for allowing us to
publish his pictures of Budapest on the website.
We benefited from material and advice kindly given by last year's
program chairs Michael Gelfond and John Gallagher. Thank you very much!

On behalf of the whole LP community, we would like to thank all
authors who have submitted a paper, the 41 members of the program
committee: Elvira Albert (U.C. Madrid), Sergio Antoy (Portland State
Univ.), Marcello Balduccini (Kodak Research Laboratories), Manuel
Carro (Technical University of Madrid (UPM)), Michael Codish (Ben
Gurion Univ.), Veronica Dahl (Simon Fraser Univ.), Marina De Vos
(Univ. of Bath), Alessandro Dal Pal\`u (Universita degli Studi di
Parma), Bart Demoen (K.U. Leuven), Thomas Eiter (T.U. Wien), Esra
Erdem (Sabanci University), Thom Fr\"uhwirth (Univ. of Ulm), Andrea
Formisano (Univ. of Perugia), Maria Garcia de la Banda (Monash Univ.),
Marco Gavanelli (University of Ferrara), Hai-Feng Guo (Univ. of
Nebraska, Omaha), Gopal Gupta (Univ. of Texas, Dallas), Katsumi Inoue
(National Inst. of Informatics, Japan), Angelika Kimmig (K.U. Leuven),
Joohyung Lee (Arizona State University), Evelina Lamma (Univ. of
Ferrara), Nicola Leone (University of Calabria), Yuliya Lierler
(Univ. of Kentucky), Boon Thau Loo (Univ. of Pennsylvania), Michael
Maher (R.R.I., Sydney), Alessandra Mileo (DERI Galway), Jose Morales
(U.P. Madrid), Enrico Pontelli (New Mexico State Univ.), Gianfranco
Rossi (Univ. of Parma), Beata Sarna-Starosta (Cambian, Vancouver),
Torsten Schaub (Univ. of Potsdam), Tom Schrijvers (Universiteit Gent),
Fernando Silva (Univ. of Porto), Tran Cao Son (New Mexico State
University), Terrance Swift (Univ. Nova de Lisboa), P\'eter Szeredi
(Budapest Univ. of Technology and Economics), Francesca Toni (Imperial
College London), Mirek Truszczynski (University of Kentucky), Germ\'an
Vidal (U.P. of Valencia), Stefan Woltran (Vienna University of
Technology), and Neng-Fa Zhou (CUNY, New York).

A particular thanks goes to the 96 external referees, namely:
Alicia Villanueva,
Amira Zaki,
Ana Paula Tom\'as,
Andrea Bracciali,	
Antonis Bikakis,
Antonis Kakas,
Brian Devries,
C.\ R.\ Ramakrishnan,
Chiaki Sakama,
Christoph Redl,
Christopher Mears,
Dale Miller,
Daniel De Schreye,
Daniela Inclezan,	
David Brown,
Demis Ballis,
Dimitar Shterionov,
Dragan Ivanovic,
Evgenia Ternovska,	
Fabio Fioravanti,	
Fabrizio Riguzzi,
Fangkai Yang,
Fausto Spoto,
Feliks Klu\'{z}niak,
Francesco Calimeri,
Francesco Ricca,
Fred Mesnard,
Gianluigi Greco,
Giovanni Grasso,
Gregory Duck,
Gregory Gelfond,
In\^es Dutra,
Jesus M. Almendros-Jimenez,
Joost Vennekens,
Juan Manuel Crespo,
Julio Mari\~no,
Kyle Marple,
Marco Alberti,
Marco Maratea,
Mario Alviano,
M\'ario Florido,
Marius Schneider,	
Martin Gebser,
Masakazu Ishihata,
Massimiliano Cattafi,
Matthias Knorr,
Maurice Bruynooghe,
Max Ostrowski,
Michael Bartholomew,
Michael Hanus,
Michael Morak,
Minh Dao-Tran,
Mutsunori Banbara,
Naoki Nishida,
Naoyuki Tamura,
Neda Saeedloei,
Nicola Capuano,
Nicolas Schwind,
Noson Yanofsky,
Nysret Musliu,
Orkunt Sabuncu,
Pablo Chico De Guzm\'an
Paolo Torroni,
Paul Tarau,
Peter James Stuckey,
Peter Sch\"uller,
Philipp Obermeier,
Puri Arenas-Sanchez,
R\'emy Haemmerl\'e,
Rafael Del Vado Virsela,
Ricardo Rocha,
Richard Min,
Robert Craven,
Roland Kaminski,
Samir Genaim,
Sandeep Chintabathina,
Santiago Escobar,
Sara Girotto,
Sean Policarpio,
Simona Perri,
Slim Abdennadher,	
Sofia Gomes,
Stefania Costantini,
Stefano Bistarelli,
Thomas Krennwallner,
Thomas Str\"{o}der,
Tomoya Tanjo,
Torben Mogensen,
Umut Oztok,
Valerio Senni,
Victor Marek,
Victor Pablos Ceruelo,
Wolfgang Dvo\v{r}\'ak,
Wolfgang Faber,
Yana Todorova, and
Yunsong Meng.

\medskip

Throughout this period, we could always rely on ALP. Our gratitude
goes to the ALP president Gopal Gupta, to the Conference chair Manuel
(Manolo) Carro, and to all the ALP Executive committe members.  We
already thanked the invited speakers and the tutorialist above, but we
would like to stress here our thank to them.  David Tranah (Cambridge)
and Ilkka Niemel\"a (TPLP editor) deserve our thanks for their
kindness and their precious support in all publication
stages. Similarly, a thanks goes to Marc Herbstritt from Dagstuhl,
Leibniz Center for Informatics, for the support in publication of the
Technical Communication.

\medskip

Our thanks also go to the the sponsors of the meeting, namely the
Association for Logic Programming (ALP), the Artificial Intelligence
Section of the John von Neumann Computer Society,  the Aquincum
Institute of Technology (AIT) of Budapest, Alerant System Inc, and Google
(female researchers grant).
Finally, a well-deserved
thank you goes to Easychair developers and managers. This amazing free
software allowed us to save days of low level activities. Similarly,
the joint work of the two co-chairs would have been extremely more
difficult and expensive without the Dropbox and Skype services.

\medskip

\hfill\begin{tabular}{r}
September 2012\\
Agostino Dovier and V\'{\i}tor Santos Costa\\
Program Committee Chairs and Guest Editors
\end{tabular}

\pagebreak

This electronic appendix reports the contents of the Volume 17, Issue 1
(DOI 10.4230/LIPIcs.ICLP.2012.42)
of the Leibniz International Proceedings in Informatics (LIPIcs) series,
published on-line through the Dagstuhl Research Online Publication
Server (DROPS) publishing position papers, technical communications,
and doctoral consortium contributed to the 28th International
Conference on Logic Programming (ICLP 2012).

\section*{Position Paper}

\tocTitle{Simulation Unification: Beyond Querying Semistructured Data}{00}
\tocAuthors{Fran\c{c}ois Bry, Sebastian Schaffert}

\section*{Technical Communication}

\tocTitle{A Logic Programming approach for Access Control over RDF}{00}
\tocAuthors{Nuno Lopes, Sabrina Kirrane, Antoine Zimmermann, Axel
  Polleres and Alessandra Mileo}

\tocTitle{Modeling Machine Learning and Data Mining Problems
    with FO($\cdot $)}{00}
\tocAuthors{Hendrik Blockeel, Bart Bogaerts, Maurice Bruynooghe, Broes De Cat,
    Stef De Pooter, Marc Denecker, Anthony Labarre, Jan Ramon and
    Sicco Verwer.}

\tocTitle{Paving the Way for Temporal Grounding}{00}
\tocAuthors{Felicidad Aguado, Pedro Cabalar, Mart\'{\i}n Di\'eguez, Gilberto P\'erez and Concepcion Vidal.}

\tocTitle{Preprocessing of Complex Non-Ground Rules in Answer Set Programming}{00}
\tocAuthors{Michael Morak and Stefan Woltran.}

\tocTitle{Two-Valued Logic Programs}{00}
\tocAuthors{Vladimir Lifschitz.}

\tocTitle{LOG-IDEAH: ASP for Architectonic Asset Preservation}{00}
\tocAuthors{Marina De Vos, Julian Padget, Vivana Novelli and Dina D'Ayala.}

\tocTitle{aspeed: ASP-based Solver Scheduling}{00}
\tocAuthors{Holger Hoos, Roland Kaminski, Torsten Schaub and Marius Schneider.}

\tocTitle{Static Type Inference for the Q language using Constraint Logic Programming}{00}
\tocAuthors{Zsolt Zombori, J\'anos Csorba and P\'eter Szeredi.}

\tocTitle{ASP at Work: An ASP Implementation of PhyloWS}{00}
\tocAuthors{Enrico Pontelli, Tiep Le, Hieu Nguyen and Tran Cao Son.}

\tocTitle{Improving Quality and Efficiency in Home Health Care: an application of Constraint Logic Programming for the Ferrara NHS unit}{00}
\tocAuthors{Massimiliano Cattafi, Rosa Herrero, Marco Gavanelli, Maddalena Nonato and Juan Jos\'e Ramos Gonzalez.}

\tocTitle{An FLP-Style Answer-Set Semantics for Abstract-Constraint Programs with Disjunctions}{00}
\tocAuthors{Johannes Oetsch, J\"org P\"uhrer and Hans Tompits.}

\tocTitle{A Tarskian Semantics for Answer Set Programming}{00}
\tocAuthors{Marc Denecker, Yuliya Lierler, Mirek Truszczynski and Joost Vennekens.}

\tocTitle{Lazy model expansion by incremental grounding}{00}
\tocAuthors{Broes De Cat, Marc Denecker and Peter Stuckey.}

\tocTitle{Logic + control: An example}{00}
\tocAuthors{W\l odek Drabent.}

\tocTitle{Tabling for infinite probability computation}{00}
\tocAuthors{Taisuke Sato and Philipp Meyer.}

\tocTitle{Improving Lazy Non-Deterministic Computations by Demand Analysis}{00}
\tocAuthors{Michael Hanus.}

\tocTitle{Surviving Solver Sensitivity: An ASP Practitioner's Guide}{00}
\tocAuthors{Bryan Silverthorn, Yuliya Lierler and Marius Schneider.}

\tocTitle{On the Termination of Logic Programs with Function Symbols}{00}
\tocAuthors{Sergio Greco, Francesca Spezzano and Irina Trubitsyna.}

\tocTitle{Towards Testing Concurrent Objects in CLP}{00}
\tocAuthors{Elvira Albert, Puri Arenas and Miguel Gomez-Zamalloa.}

\tocTitle{Generating Event-Sequence Test Cases by Answer SetProgramming with the Incidence Matrix}{00}
\tocAuthors{Mutsunori Banbara, Naoyuki Tamura and Katsumi Inoue.}

\tocTitle{Answer Set Solving with Lazy Nogood Generation}{00}
\tocAuthors{Christian Drescher and Toby Walsh.}

\tocTitle{Reconciling Well-Founded Semantics of DL-Programs and AggregatePrograms}{00}
\tocAuthors{Jia-Huai You, John Morris and Yi Bi.}

\tocTitle{Stable Models of Formulas with Generalized Quantifiers}{00}
\tocAuthors{Joohyung Lee and Yunsong Meng.}

\tocTitle{Extending C+ with Composite Actions for Robotic Task Planning}{00}
\tocAuthors{Xiaoping Chen, Guoqiang Jin and Fangkai Yang.}

\tocTitle{Deriving a Fast Inverse of the Generalized Cantor N-tupling Bijection}{00}
\tocAuthors{Paul Tarau.}

\tocTitle{Unsatisfiability-based optimization in clasp}{00}
\tocAuthors{Benjamin Andres, Benjamin Kaufmann, Oliver Matheis and Torsten Schaub.}

\tocTitle{Visualization of Constraint Handling Rules through Source-to-Source Transformation}{00}
\tocAuthors{Slim Abdennadher and Nada Sharaf.}

\tocTitle{Applying Machine Learning Techniques to ASP Solving}{00}
\tocAuthors{Marco Maratea, Luca Pulina and Francesco Ricca.}

\tocTitle{Flexible Solvers for Finite Arithmetic Circuits}{00}
\tocAuthors{Nathaniel Filardo and Jason Eisner.}

\tocTitle{The additional difficulties for the automatic synthesis of specifications posed by logic features in functional-logic languages}{00}
\tocAuthors{Giovanni Bacci, Marco Comini, Marco A. Feli\'u and Alicia Villanueva.}

\tocTitle{Answering Why and How questions with respect to a frame-based knowledge base: a preliminary report}{00}
\tocAuthors{Shanshan Liang, Nguyen Ha Vo and Chitta Baral.}

\tocTitle{Logic Programming in Tabular Allegories}{00}
\tocAuthors{Emilio Jes\'us Gallego Arias and James B. Lipton.}

\tocTitle{Possibilistic Nested Logic Programs}{00}
\tocAuthors{Juan Carlos Nieves and Helena Lindgren.}

\tocTitle{Using Answer Set Programming in the Development of Verified Software}{00}
\tocAuthors{Florian Schanda and Martin Brain.}

\tocTitle{An Answer Set Solver for non-Herbrand Programs: Progress Report}{00}
\tocAuthors{Marcello Balduccini.}

\tocTitle{CHR for Social Responsibility}{00}
\tocAuthors{Veronica Dahl, Bradley Coleman, Emilio Miralles and Erez Maharshak.}

\tocTitle{A Concurrent Operational Semantics for Constraint Functional Logic Programming}{00}
\tocAuthors{Rafael Del Vado V\'{\i}rseda, Fernando P\'erez Morente and Marcos Miguel Garc\'{\i}a Toledo.}

\section*{Doctoral Consortium Contributions}

\tocTitle{Software Model Checking by Program Specialization}{00}
\tocAuthors{Emanuele De Angelis}

\tocTitle{Temporal Answer Set Programming}{00}
\tocAuthors{Mart\'in Di\'eguez}

\tocTitle{A Gradual Polymorphic Type System with Subtyping for Prolog}{00}
\tocAuthors{Spyros Hadjichristodoulou}

\tocTitle{ASP modulo CSP: The clingcon system}{00}
\tocAuthors{Max Ostrowski}

\tocTitle{An ASP Approach for the Optimal Placement of the Isolation Valves in a Water Distribution System}{00}
\tocAuthors{Andrea Peano}

\tocTitle{Answer Set Programming with External Sources}{00}
\tocAuthors{Christoph Redl}

\tocTitle{Together, Is Anything Possible? A Look at Collective Commitments for Agents}{00}
\tocAuthors{Ben Wright}

\end{document}